# Microdroplet self-propulsion during dropwise condensation on lubricant-infused surfaces

Jianxing Sun[a], Patricia Weisensee[a,b,*]

[a] Department of Mechanical Engineering & Materials Science, Washington University in St. Louis, MO

[b] Institute of Materials Science and Engineering, Washington University in St. Louis, USA

* corresponding author: p.weisensee@wustl.edu

# Abstract

Water vapor condensation is common in nature and widely used in industrial applications, including water harvesting, power generation, and desalination. As compared to traditional filmwise condensation, dropwise condensation on lubricant-infused surfaces (LIS) can lead to an order-of-magnitude increase in heat transfer rates. Small droplets ($D \leq 100$ μm) account for nearly 85 % of the total heat transfer and droplet sweeping plays a crucial role in clearing nucleation sites, allowing for frequent re-nucleation. Here, we focus on the dynamic interplay of microdroplets with the thin lubricant film during water vapor condensation on LIS. Coupling high-speed imaging, optical microscopy, and interferometry, we show that the initially uniform lubricant film re-distributes during condensation. Governed by lubricant height gradients, microdroplets as small as 2 μm in diameter undergo rigorous and gravity-independent self-propulsion, travelling distances multiples of their diameters at velocities up to 1100 μm/s. Although macroscopically the movement appears to be random, we show that on a microscopic level capillary attraction due to asymmetrical lubricant menisci causes this gravity-independent droplet motion. Based on a lateral force balance analysis, we quantitatively find that the sliding velocity initially increases during movement, but decreases sharply at shorter inter-droplet spacing. The maximum sliding velocity is inversely proportional to the oil viscosity and is strongly dependent of the droplet size, which is in excellent agreement with the experimental observations. This novel and non-traditional droplet movement is expected to significantly enhance the sweeping efficiency during dropwise condensation, leading to higher nucleation and heat transfer rates.

# Introduction

Vapor condensation is ubiquitous in nature[1,2] and at the basis of a wide range of industrial applications, such as power generation[3], air conditioning[4], water harvesting[5], and desalination[6]. Filmwise condensation, in which the condensate forms a thick liquid film, is common, owing to the high wettability of most solid surfaces. This film acts as a thermal barrier to heat transfer.[7] In contrast, dropwise condensation, in which the condensate forms discrete droplets that grow through direct condensation, coalescence, and gravity-assisted shedding, leads to an order-of-magnitude improvement in heat transfer performance compared to the filmwise mode.[8–10] Lubricant-infused surfaces (LIS) promote dropwise condensation and, unlike nanostructured superhydrophobic surfaces, maintain stable operating conditions even at high supersaturation.[11–13] Droplets on LIS have extremely low contact angle hysteresis due to the presence of a lubricating film between working fluid and substrate,[14] which results in a remarkable mobility of the condensed droplets.[15] When droplets rest on an oil film, a meniscus forms around the base of the droplet. Additionally, many lubricants cloak the working fluid, *i.e.*, a thin (10s of nm) oil layer engulfs the droplet.[16,17] The meniscus and cloak lead to increased viscous dissipation and lower droplet sliding velocities.[18–24] On the other hand, overlapping menisci can also induce a net attractive force between droplets,[25–28] potentially enhancing gravity-induced droplet sweeping. Frequent droplet sweeping is essential to remove condensate, to renew nucleation sites, and to maintain high heat transfer rates during condensation. It also plays a key role in coalescence-induced droplet growth and can accelerate oil drainage, causing LIS to fail. Hence, droplet movement during condensation on LIS is of great importance to ensure an efficient and long-term stable operation. As most previous studies have quantified the dynamics of millimetric droplets, there continues to be a lack of experimental characterization on the mobility of microscopic droplets on LIS, despite their importance for heat transfer; microdroplets with diameters $D \leq 100$ μm account for nearly 85 % of the total heat transfer.[8,29] Due to an increased meniscus-to-droplet volume ratio for microdroplets, we expect a strong relationship between the dynamics of the lubricating oil film and droplet movement.

In this paper, we reveal rigorous and gravity-independent self-propulsion of microscopic droplets as small as 2 μm in diameter, which traverse distances multiples of their diameters at velocities up to 1100 μm/s. Gravity-independent droplet sliding has previously been reported only for droplets larger than 100 μm in diameter, and no explanation on its nature has been provided.[15] We hypothesize that this microdroplet mobility, which cannot be observed on solid (super)hydrophobic surfaces, stems from a dynamic re-distribution of the lubricant film, and has a great potential to increase sweeping and coalescence rates during condensation. Furthermore, we propose that this capillary-driven droplet mobility enables the operation of condensers in space or in mobile application, where droplet removal cannot rely on gravity-assisted shedding. Here, we use high-speed interferometry to unveil the transient re-distribution of the oil film during condensation caused by the development of oil wetting ridges. We distinguish two separate regions: oil-rich and oil-poor regions. We show that the lubricant film re-distribution and the associated oil regions cause two distinct condensate droplet movement regimes, where we observe the self-propulsion of microdroplets at the boundary between oil-rich and oil-poor regions. Through experiments and modeling, we further elucidate the dependence of velocity and displacement on oil viscosity, droplet size, inter-droplet spacing, and initial oil film thickness.

## Materials and methods

**Preparation of lubricant-infused surfaces (LIS)**

To fabricate the substrates, a plain microscope glass slide (Thermo Scientific) was first cleaned with acetone, isopropanol, and de-ionized (DI) water, and dried with compressed $N_2$. Subsequently, the glass slide was treated with a layer of commercially available superhydrophobic agent (Glaco Mirror Coat, soft 99 Co.), which is an alcohol-based suspension of silica nanoparticles.[30] Then, the sample was placed in the fume hood for one hour at room temperature to ensure complete solvent evaporation. As shown in the inset SEM images in Fig. 1, a layer of relatively uniform glaco coating layer is obtained. Since the glaco coating will be infused with a micrometric oil film, the small height gradients are not expected to affect the overall performance. To obtain the lubricant-infused surface, the nanostructured surface was impregnated with Krytox oils *via* spin coating at 1000 – 3000 rpm, depending on the oil viscosity. The initial oil thickness ranged from 4 to 15 µm, as determined by interferometry (see **Fig. S1** in the Supplement Information (SI)). We measured contact angles of DI water on the Glaco-coated glass slide sample of 165 ± 3° before oil impregnation. Once infused the apparent contact angle decreases, but the samples retain their extremely small roll-off angles and no pinning of millimetric water droplets was observed. Krytox GPL series oils all have similar surface tensions in air, $\gamma_{oa}$ = 17 ± 1 mN/m and in water $\gamma_{ow}$ = 53 mN/m,[24] but very different viscosities, which are listed in Table 1. DI water ($\gamma_{wa}$ = 72.4 mN/m) is used as working liquid in this work. The oils will cloak condensed water droplets due to a positive spreading coefficient, $S_{ow(a)}$ = $\gamma_{wa}$ − $\gamma_{oa}$ − $\gamma_{ow}$ = 2.4 mN/m > 0.[16,17,21] Both Krytox oils and silicone-based vacuum pump oil are widely used in design of stable LISs in previous studies[24,31] and completely spread on the nanostructure-textured surfaces in the presence of both of air and water[19,32,33], indicating the formation of a thin oil layer between water droplets and the substrate. Using 3D confocal fluorescence microscopy we confirm the presence of the thin film underneath the droplets (see **Fig. S2**).

Table 1: Viscosity of the lubricants at room temperature

| Lubricant | Viscosity $\mu_o$ [cP] |
|---|---|
| Krytox GPL 102 | 73 |
| Krytox GPL 104 | 350 |
| Krytox GPL 105 | 1034 |
| Krytox GPL 106 | 1627 |
| Vacuum pump oil | 500 |
| DI water | 1 |

**Experimental setup**

To observe the dynamics of the lubricant oil and droplets during condensation, the sample was mounted on a Linkam PE120 cold stage using double-sided thermally conductive aluminum-based tape (Marian Inc., Chicago) and placed under an upright microscope (Nikon Eclipse LV100). The thermo-electric cold stage was set to 2 °C during all experiments and the temperature on the sample surface ranged from 5 to 8 °C, as measured by a surface-mounted platinum RTD (Omega Engineering Inc.). A well-sealed Erlenmeyer flask containing DI water was heated on a hot plate to approximately 50 °C, as measured by a digital thermocouple meter (Digi-Sense). Compressed $N_2$ was supplied to the bottom of the container at ≈ 10 liter per minute (LPM) using a flexible PVC tube. The vapor-saturated $N_2$ was then guided through a second PVC tube to the sample to achieve 100 % relative humidity near the sample surface. The flow rate was low enough such that it had negligible influence on the movement of the condensed droplets. The experiments were performed in an open environment (room temperature ≈ 20 °C at 30 – 40 % relative humidity). The condensation experiments were conducted on horizontally placed samples. Between individual experimental runs, dry $N_2$ was gently blown over the sample to prevent flooding. The experimental setup, as shown in Fig. 1, was placed on an optical table (TMC) to isolate the setup from environmental vibrations. Two brightfield objectives (50× L Plan SLWD, Nikon, and 100× T Plan Fluor EPI SLWD, Nikon) were used to visualize droplet sliding. Two interferometry objective lenses (10× and 50× CF IC Epi Plan DI Interferometry Objectives, Nikon) were used to visualize transient oil thickness gradients.

Videos were recorded using either an AmScope microscope color camera or a Photron FASTCAM Mini AX200 high-speed camera at 30 to 10,000 frames per second (fps).

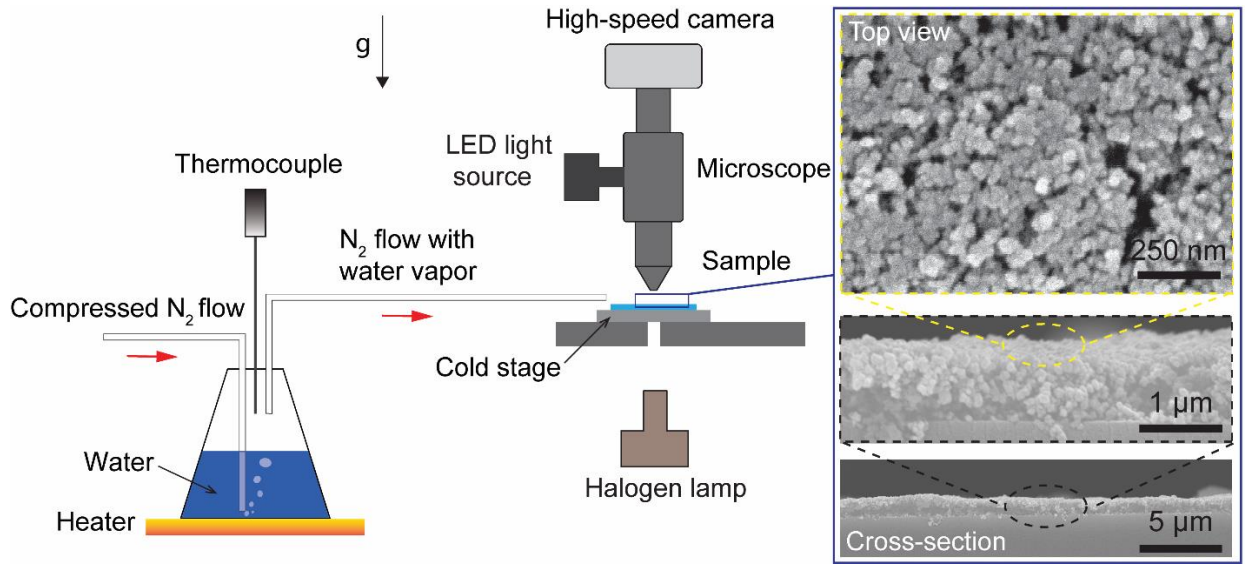

Fig. 1. Schematic of the experimental setup. (Inset) Cross-sectional and top view scanning electron microscope (SEM) images of the nanoparticle coating

**Interferometry**

Interference microscopy allows for non-contact profiling of the dynamic re-distribution of the oil film on LIS. We used monochromatic light sources (SOLIS-623C Red LED and Green LED M530L3, Thorlabs) to measure the oil meniscus shape surrounding the droplets and a white light halogen lamp to qualitatively visualize the dynamic change in oil film thickness during condensation. Since the spatial resolution of the Photron high-speed camera is limited to 370 nm/pixel at 50× magnification, the maximum meniscus slope $\alpha$ that could be discerned with the green LED ($\lambda$ = 530 nm) was approximately 35°. The change in oil meniscus height $\Delta h$ between two neighboring interference fringes (bright and dark) of distance $\Delta x$ is then given by[34]:

$$\Delta h = \frac{\lambda}{2} = \Delta x \tan \alpha \ . \tag{1}$$

## Results and discussion

**Oil meniscus shape surrounding a microdroplet on LIS**

The oil meniscus surrounding a droplet becomes increasingly important as the droplet size decreases, as it plays a key role in droplet dynamics on LIS.[31] The wetting state of a millimetric droplet placed on LIS has been extensively studied.[17,19,20,23] As shown in Fig. 2a, an annular oil ridge (*i.e.*, meniscus) surrounds the base of the droplet and a thin oil film separates the droplet from the substrate. Here, we deposited a sessile droplet on the surface using a 33 gauge needle and let it evaporate naturally at room temperature to obtain a droplet of the desired size (~ 100s μm). Microdroplets have a similar wetting state, however, the meniscus-to-droplet size ratio is significantly increased. Fig. 2b shows a condensed microdroplet in side view (20 x objective lens), clearly showing a very pronounced meniscus which is of comparable size to the droplet. To quantify the relationship between the meniscus shape and the droplet size at the microscale, we experimentally observe the shape change of the oil meniscus as a condensed water droplet evaporates on LIS (the evaporation of Krytox oil is negligible). By controlling the cold stage temperature, we first let the droplet slowly evaporate in nearly constant contact angle mode.[35] This slow evaporation is critical to obtain quasi-steady droplet and meniscus shapes, *i.e.*, to give the meniscus sufficient time to relax to an equilibrium position before image acquisition. Figure 2c shows the evolution of the profiles of the microdroplets and corresponding oil menisci during evaporation (for comparison, see Figs. 2a,b). Interestingly, the maximum meniscus height remains nearly constant when the droplet diameter is greater than $D \geq 130$ μm, likely caused by the limited oil supply on LIS. For droplets $D < 130$ μm, the maximum meniscus height decreases with decreasing droplet size.

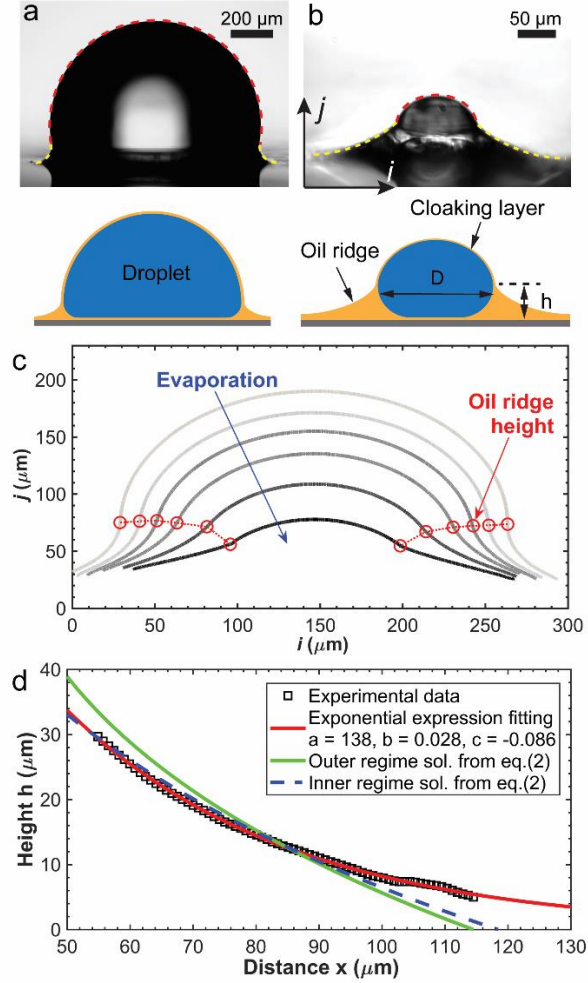

Fig 2. Wetting states and meniscus profiles for droplets on LIS. Images and schematics of (a) millimetric and (b) microscale droplets on LIS. (c) Evoltion of profiles of droplet and oil ridge during evaporation on LIS. The red symbols at the inflection points between droplet and meniscus profiles represent the maximum oil ridge height. The corresponding coordinate system is shown in (b). (d) Lubricant meniscus profile surrounding a water droplet with $D$ = 110 µm (similar to Fig. 2b) on LIS fitted with different models. Black squares: Experimenal data for the meniscus height as a function of the distance $x$ from the droplet center. Green and dotted blue: First order analytical approximations for eq. (2).[23] Red: Exponential expression with the fitting parameters, a =138, b = 0.028 and c = -0.086.

To allow for mathematical modeling (see modeling section below), we are interested in obtaining an analytical expression describing the meniscus shape. For an isolated object placed at the liquid-air interface of a bulk liquid, the interface will deform to form a well-developed meniscus whose shape is governed by the Laplace equation:[23]

$$\gamma_{la} \left\{ \frac{d^2h/dx^2}{[1+(dh/dx)^2]^{3/2}} + \frac{dh/dx}{x[1+(dh/dx)^2]^{1/2}} \right\} = g(\rho_l - \rho_a)h, \qquad (2)$$

where $x$ denotes the distance from the center of the object; $h = h(x)$ is the height of the meniscus above the equilibrium liquid level, and $\rho_l$ and $\rho_a$ are the liquid and air densities, respectively. For eq. (2), a general analytical solution is not available. However, for the special case of a symmetrical object and unlimited liquid supply, eq. (2) can be approximated by two asymptotic regimes at $x \ll l_c$ ('outer' regime) and $x \gg l_c$ ('inner' regime), where $l_c = (\gamma_{oa}/\rho_o g)^{1/2}$ is the capillary length.[23,36] Because $x \ll l_c$ in the 'outer' regime, the hydrostatic pressure on the right hand of eq. (2) becomes negligible.[37] For Krytox oil, the capillary length is approximately $l_c \approx 1$ mm. Furthermore, capillary forces dominate over gravitational forces, given a Bond number $Bo = \Delta \rho g D^2/\gamma_{wo} \approx 0.00138 \ll 1$. Unfortunately, both of the 'inner' and 'outer' analytical solutions are not a good fit to represent the meniscus around a microdroplet on LIS and fitting parameters are still necessary to be found from matching the analytical models to experimental data, as shown in Fig. 2c. However, we found that the meniscus profile can be well described by an exponential expression: $h(x) = a \exp(-bx) + c$,[38] where the fitting parameters $a$, $b$ and $c$ are determined empirically. This exponential approximation will be used for the remainder of this study. We found this exponential expression to be a good representation of the oil meniscus shape for the majority of microscale droplets, using different fitting parameters $a$ and $b$. As suggested previously for larger objects at the surface of an infinite bath,[17] we also observe a dependence of the droplet size on these two parameters. As mentioned above, and shown in Fig. 2c, the meniscus height saturates above a critical droplet size ($D \approx 130$ μm) due to the scarcity of available lubricant on the surface. A more rigorous characterization of the physical relationship between object shape, size, and lubricant availability is beyond the scope of this work, but should be addressed in a future study.

**Transient oil re-distribution and formation of oil menisci**

The surface of a pristine sample is covered with an oil film of uniform thickness. During condensation, droplets nucleate at the oil-air interface and submerge into the oil due to capillary forces and cloaking.[16] When the condensed droplets are still sufficiently small, we observe dense microscopic droplets ($D \approx 3$ μm < oil film thickness) submerged within the oil, having a relatively smooth oil-air interface. As droplets continue to grow *via* direct condensation, the oil between submerged droplets drains, allowing for droplet coalescence. The characteristic time for droplet coalescence is directly proportional to the oil viscosity.[24,39] Hence, immersed droplets grow faster for the low viscosity oil (GPL 102, 73 cP) and the droplet size distribution becomes less uniform, as shown in Fig. 3a. A previous study using silicone oil of varying viscosity reported similar observations, where the polydispersity in droplet size distribution on a 100 cP surface was triple that of a 1000 cP surface after 250 seconds.[16] After protruding the oil-air interface, droplets move around on the sample and coalesce with neighboring droplets. We observe the projected droplet shapes are almost perfectly circular throughout on GPL 102-impregnated surfaces. For the high viscosity oil (GPL 106, 1627 cP), the coalescence resistance is higher and the densely packed pattern of uniformly sized microdroplets persists for longer periods of time, as shown in Fig. 3b.[16,24] Eventually, coalescence is promoted due to a higher squeezing pressure as droplets grow bigger. Immediately following coalescence, droplet-vacant regions appear. As condensation continues, larger droplets with smaller ones trapped in their oil menisci form clusters due to capillary interactions. Relatively larger droplets ($D$ > 90 µm) of comparable size typically remain in this cluster without coalescing for a period on the order of 10s of seconds. The droplets in these clusters become distorted from circular to polyhedral due to energy and area minimization.[16] In the brightfield-videos, darkened regions appear surrounding these droplet clusters, as seen in Figs. 3a, b. We attribute this darkened region to the existence of menisci, *i.e.*, oil-rich regions surrounding large droplets or droplet clusters. The total area of such oil-rich regions depends strongly on the sizes of the droplets and the initial lubricant layer thickness. Due to conservation of mass, oil drains from the peripheral areas as it accumulates near the droplet clusters, and oil-poor regions form. The location and area-ratio of oil-rich and oil-poor regions change dynamically due to droplet growth and movement during condensation. For example, oil-rich regions can re-develop within oil-poor regions due to the same re-distribution mechanism after droplet nucleation and growth. Over time, the oil thickness in the oil-rich and oil-poor regions decreases because of continuous lubricant depletion.[21] Only once the lubricant is severely depleted, this dynamic re-distribution no longer occurs. Fig. 4 visualizes the uneven distribution of lubricant after the oil re-distribution process using interferometry. Using 3D confocal fluorescence microscopy, we measured the oil film thickness of the oil-poor regions after two condensation sweeping cycles to ≈ 3 ± 0.3 µm. In the oil-rich region surrounding a droplet with $D$ ≈ 37 µm, the meniscus is up to 12 ± 0.3 µm tall. Due to the incompatibility of Krytox oils with common fluorescent dyes, we used vacuum pump oil (500 cp) dyed with Lumogen F Red 305 (BASF Colors & Effects) for these confocal microscopy measurements (see **Fig. S2**).

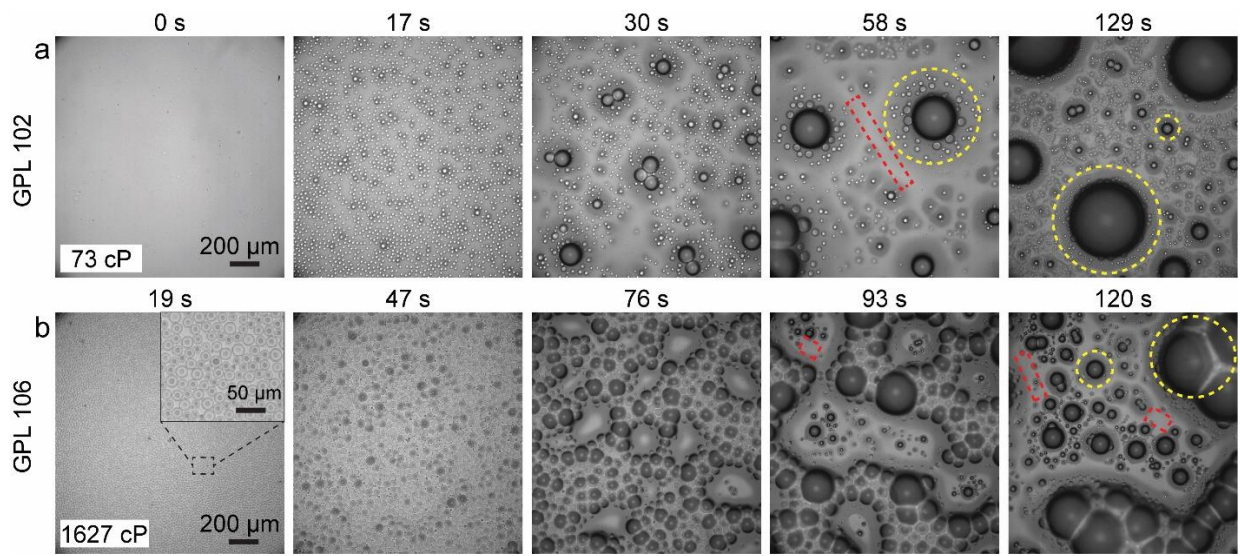

Fig. 3. Time sequence showing the re-distribution of the lubricant layer for (a) Krytox GPL 102 and (b) Krytox GPL 106 during early stages of water vapor condensation. Examples of oil-rich and oil-poor regions are outlined by dashed yellow circles and red rectangles, respectively.

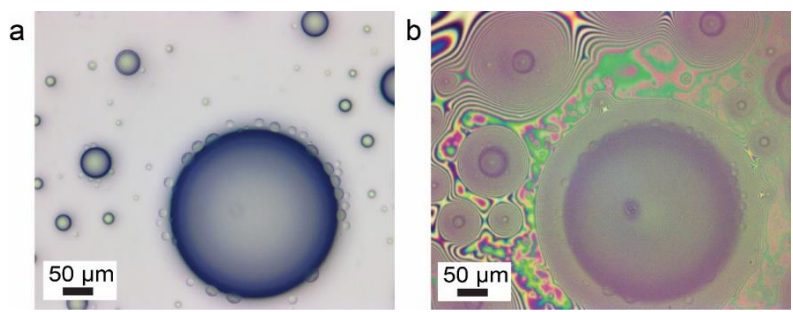

Fig. 4. Visualization of the uneven lubricant layer after the initial oil re-distribution in the same region on a sample using (a) brightfield and (b) interferometry imaging.

**Movement of microdroplets induced by asymmetrical oil menisci**

The coexistence of oil-rich and oil-poor regions, which prevail at varying length scales after the initial re-distribution, strongly influences the mobility of microdroplets (see **Movie S1**). As shown in Fig. 5a, nucleation is predominantly found in the oil-poor regions (dashed red rectangle), since the oil film is thin and poses a low conduction resistance (*i.e.*, the oil-air interfacial temperature is lowest).[16,22,29] The microdroplets are generally stationary before growing large enough through direct condensation, at which point they suddenly coalesce with nearby droplets. Their growth and mobility are similar to those of droplets condensing on (super)hydrophobic surfaces, but will not jump out of plane because of strong adhesion and viscous dissipation.[40,41] Conversely, microdroplets display high mobility in the oil-rich and transition regions surrounding larger quasi-stationary droplets or droplet clusters. After nucleation and growth *via* direct condensation in the oil-poor region, a microdroplet protrudes the oil film and becomes visible at the periphery of an oil-rich region. Then, the microdroplet starts to move towards the big droplet along the center-to-center line and finally comes to a halt at a certain distance $r_c$ from the big droplet. This distance depends on the sizes of the microdroplet, the big droplet, and the meniscus. Droplets of all sizes are highly mobile on substrates infused with the low viscosity oil (Krytox GPL 102), so the overall displacement distance is short due to frequent random coalescence events. For high viscosity oils (Krytox GPL 104, 105,106), the movement of microdroplets is not random: microdroplets located in the transition regions spontaneously move long distances (multiples of their diameters) towards the big central droplet, sometime regardless of the existence of other neighboring droplets close-by, as shown by the colored tracks representing droplet motion in Fig. 5b (also see **Movie S3**). In Fig. 5c, we plot the displacement - velocity profiles of the moving droplets from Fig. 5b. Initially, the microdroplet accelerates and the velocity increases. After reaching a size-dependent maximum velocity $v_{max}$ at the critical distance $r_c$ from the big droplet, microdroplets decelerate. Their movement seizes as either a) the two droplets coalesce, or b) the microdroplet becomes almost fully submerged in the oil meniscus (see **Fig. S3**). The velocity profiles for different droplet sizes are self-similar, which implies that a common physical mechanism underlies the movement. Side view images in Figs. 5d exemplify a typical sequence of microdroplet mobility. We hypothesize that capillary attraction causes the observed droplet movement on LIS. When the menisci of two neighboring droplets overlap, the non-symmetry on opposing sides of a droplet induces an attractive force.[26–28] Consequently, in oil-poor regions, where menisci do not extend far enough to interact with neighboring wetting ridges, microdroplets display no or little movement.

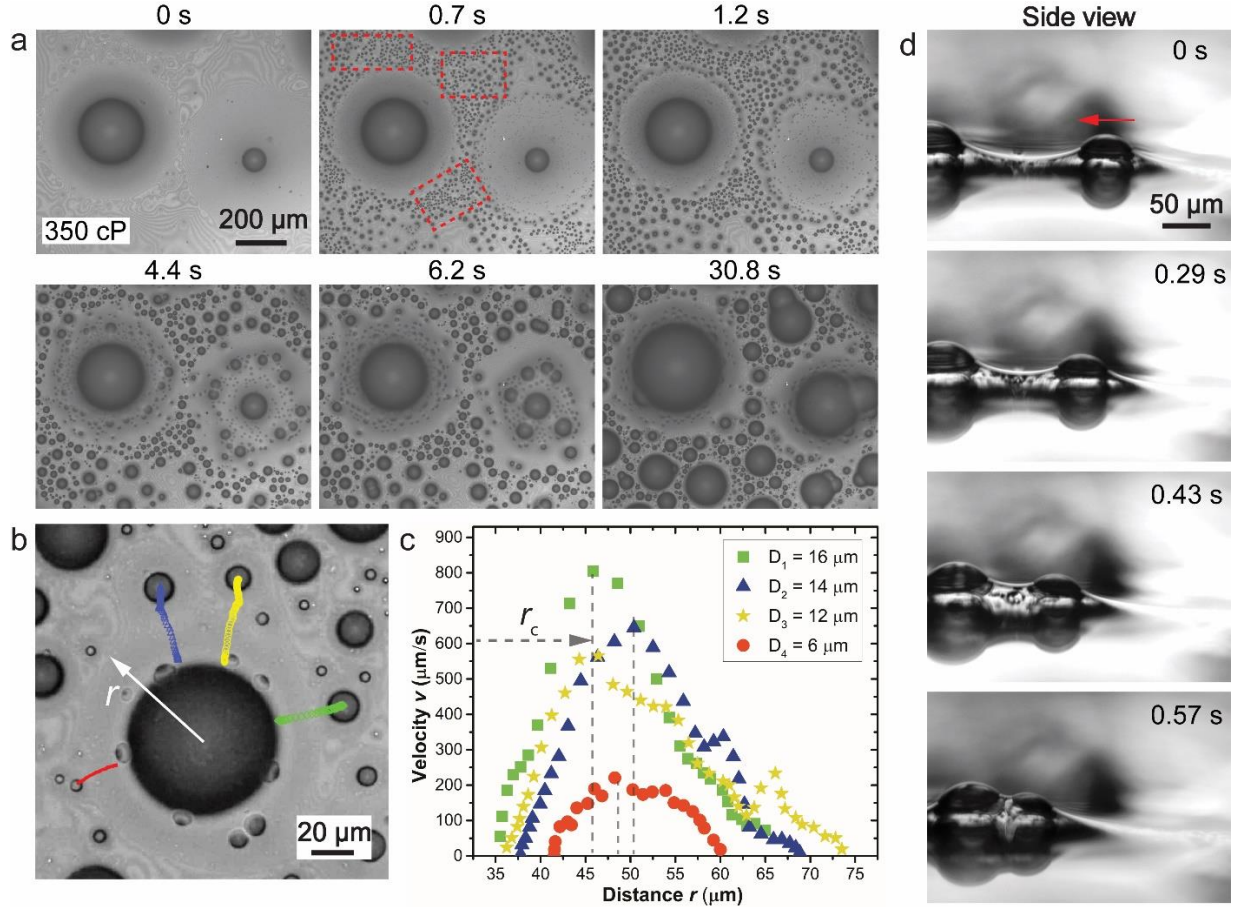

Fig. 5. Gravity-independent microdroplet movement. (a) Time sequence of a condensation process on LIS. (b) Microdroplets move toward one big pseudo-stationary droplet sitting in the center of the oil-rich region. (c) Relationship of the velocity of the small moving droplets in (b) and their displacement. (d) A typical movement in side view.

To test our hypothesis on the influence of oil menisci on the self-propulsion of droplets in oil-rich regions, we use high-speed interferometry to visualize the interplay of oil menisci and droplet movement. Fig. 6a shows typical interference fringe patterns surrounding the droplets. Following eq. (1), the change in oil thickness along the red line is plotted in Fig. 6b. We define the angle $\varphi$ as the tangent to the oil-air interface at the apparent oil-water-air triple contact line with respect to the horizontal. The angles $\varphi_A$ and $\varphi_R$ denote the advancing angle (towards the big droplet) and receding angle (away from the big droplet) for the moving microdroplet, respectively (see Fig. 6b). Due to the overlapping menisci of the two droplets, the meniscus slope $\varphi$ along the apparent fluid contact line of the microdroplet will change as the two droplets approach each other. The meniscus facing the big droplet is elevated and the slope $\varphi$ becomes smaller, *i.e.*, $\varphi_A < \varphi_R$. The horizontal component of the surface tension force along the apparent contact line becomes unbalanced and induces a lateral capillary attractive force. In most cases, the direction of movement is coincident with the maximum gradient in lubricant layer thickness, which is along the center-to-center line between two droplets.

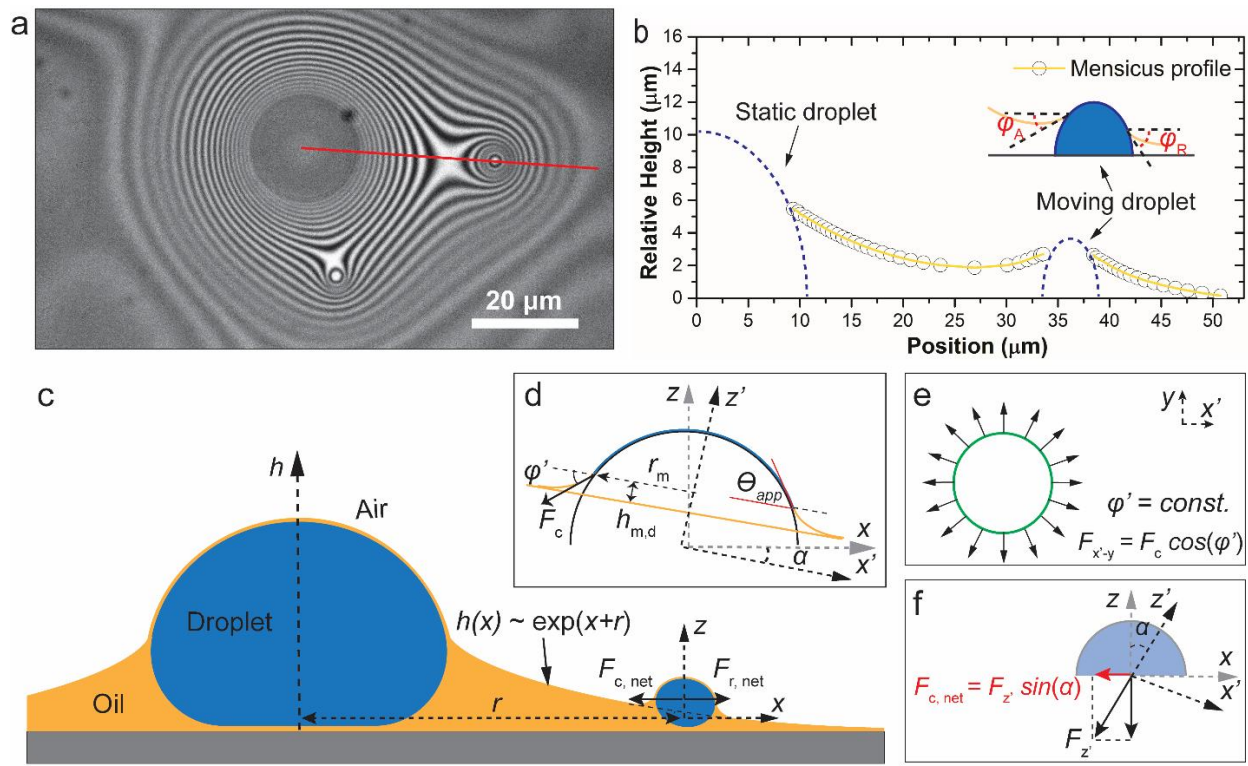

Fig. 6. Theoritical analysis of the gravity-independent droplet movement. (a) Interference patterns around microdroplets during condensation. (b) Oil film thickness change along the red line in (a). Note that the absolute oil film thickness is unknown. (c) Schematic of the proposed mechanism for capillary-induced droplet movement for a condensed droplet located in the meniscus region of a stationary larger droplet. (d) Transformation of the surface tension force from the *x-y-z* to the *x'-y-z'* coordinate system by rotation of angle *α*, where the apparent contact angle $\theta_{app}$ remains constant. (e) Balance of the componets of the capillary force in the *x'-y* coordinate system. (f) Cappillary force analysis in the x-z coordinate system. A net driving force acts in the negative x-direction.

To confirm that the observed droplet mobility is orientation- and gravity-independent, we conducted additional experiments on vertical samples and observed the same kind of movement during vapor condensation (see **Movie S2** and **Fig. S4**). To further explore whether the transient oil re-distribution and the microdroplet self-propulsion are surface-structure-dependent, we also conducted experiments on other types of samples, such as aluminum substrates with boehmite nanostructures, glass slides with Never-Wet-coating (another commercially available superhydrophobic spray) and a silicon substrate patterned with a square micropillar array (width and spacing of the micropillars: 10 µm each; height: 15 µm). These samples were overfilled with GPL 105 oil by several micrometers. A similar formation of oil-rich and oil-poor regions and similar microdroplet propulsion were observed. However, the sizes of moving droplets were generally bigger on the silicon substrate and the movement was more constrained by the microstructure. Furthermore, small droplets got trapped within the microstructures, similar to the observations by Kajiya *et al*.[22] Condensation experiments were also conducted on a smooth, silanized silion wafer coated with a thin layer of oil (which is strictly not LIS any more). The oil depleted rapidly, leading to flooding and filmwise condensation within seconds. These findings indicate that nano- or microstructures are indeed essential in retaining oil and thus guaranteeing the occurrence of transient oil re-distribution.[42,43]

Similarly, we expect there to be a minimum critical oil thickness, under which droplets no longer self-propagate, given that there is no self-propulsion without an oil layer. We conducted condensation experiment on a vertically mounted Glaco-coated glass slide sample impregnated with GPL 106 oil at 1500 rpm and monitored the droplet dynamics using a high-speed camera. When we observed a deterioration of the self-propulsion (roughly after 2.5h) using a 20x objective lens, we stopped the experiment and measured the oil film thickness using the method described in section S1 of the Supplemental Information. We repeated the experiment three times and found the average minimum oil film thickness to be around 0.6 ± 0.3 µm. Please note that the microdroplet self-propulsion still existed even after 2 hours of continuous condensation, albeit at a smaller scale. For future applications, the design of a durable LIS system is important, for example, by pairing the slippery lubricant-infused porous surface condenser with a superhydrophilic evaporator to maintain high droplet mobility throughout the lifetime of the operation cycle.[44]

**Mathematical model of the gravity-independent droplet movement**

To further reveal the underlying physical mechanism of this gravity-independent droplet movement, we developed a theoretical model based on a lateral force balance on the microdroplet. We made several simplifying assumptions: A) The model only includes a pair of droplets, a microdroplet droplet with radius $R_1$ moving towards a pseudo-stationary larger droplet with radius $R_2$, surrounded by an oil-rich region. Several concurring microdroplet movements would be considered independently of each other. B) We restrict the size ratio of the droplets to $R_1/R_2 \ll 1$ in the analysis, so that we can exclude the displacement of the bigger droplet (*i.e.*, assume it to be stationary). We assume that the larger droplet is nearly immobile unless size-comparable droplets are involved, or it approaches the capillary length of water ($l_c = (\gamma_w/\rho_w g)^{1/2} \approx 2.7$ mm), at which point gravity would dominate its movement. C) The droplet shape resembles a spherical cap with a base radius of $R_i$. When the moving droplet approaches the big droplet, the apparent contact angle will continuously change due to a changing oil film thickness. We approximate the droplets at the beginning of their movement as a hemisphere and with the mass $m_i = 2\pi\rho R_i^3/3$ with the projected droplet radius $R_i$. The mass of the two droplets is treated as constant during the studied process, as the stationary droplet has an initial large volume and the microdroplet is partially immersed in the oil, limiting direct condensation. D) The interface between the droplet, the oil cloak, and the air is represented by an effective interfacial tension $\gamma_{eff} = \gamma_{wo} + \gamma_{oa}$, which then satisfies an adapted Neumann triangle assumption in which the angles remain constant and rotate as the meniscus slope changes (see **Fig. S5**).

To determine the net force acting on the microdroplet, we set up a force balance taking into account capillary and viscous forces. As mentioned above $Bo \ll 1$, so that the gravitational force is negligible. In our experiments, the oil viscosity $\mu_o$ is much larger than that of water $\mu_w$. For an isolated millimetric/sub-millimetric droplet moving on LIS, viscous friction generally has three sources: viscous dissipation in the droplet, in the oil film underneath the droplet, and in the oil ridge surrounding the droplet.[19,45] However, as a microdroplet approaches the stationary droplet, the moving droplet will gradually immerse into the larger static oil ridge. Viscous dissipation is no longer confined to the microdroplet's own meniscus. Instead, we approximate this case with Stokes flow, i.e., the small droplet has to overcome a hydrodynamic drag force $F_{drag} = 6\pi\mu_o v R_1$. Viscous dissipation in the droplet or the thin oil film beneath the droplet are negligibly small compared to the Stokes drag. Furthermore, as the small droplet approaches the big one, oil has to squeeze out of the closing gap, which we generalize by introducing an additional distance-dependent hydrodynamic interaction resistance force: $F_{interaction} = 6\pi\mu_o v \frac{R^{*2}}{r - R_1 - R_2}$ with $R^* = R_1 R_2/(R_1 + R_2)$.[25] The closer the moving droplet gets to the stationary one the stronger this hydrodynamic interaction becomes. Overall, the total viscous resistance force can be approximately as:

$$F_{r,net} \approx 6\pi\mu_o v \left(R_1 + \frac{R^{*2}}{r - R_1 - R_2}\right), \quad (3)$$

As discussed above, overlapping menisci and different slopes attacking the droplet act as the driving force for microdroplet motion. To compute the driving force, we follow an analysis procedure as outlined in Figs. 6 d-f. First, we approximate the larger meniscus with a constant slope $\alpha(r)$, which can be obtained from the derivative of the meniscus profile $h(x) = a\exp(-bx) + c$ at $x = r$. Then, we rotate the x-z coordinate system (y axis is fixed) of the smaller droplet by the angle $\alpha$, such that the apparent contact angle $\theta_{app}$ remains constant. In the x'-y-z' coordinate system, the capillary force can be computed by integrating the surface tension exerted by the meniscus over the (apparent) lubricant-droplet-air contact line and then separately projected onto the *x'-y* plane and the *z'* axis. Forces in the *x'-y* plane are balanced, since $h_{m,d}$ = const., but there will be a net force in the negative *z'* direction, $F_z' = 2\pi r_m \gamma_o \sin(\varphi')$, where $r_m$ is the projected radius of the meniscus at the (apparent) lubricant-droplet-air contact line. For a droplet subject to a sloped meniscus, the net lateral driving force can be approximated by projecting the force $F_z'$ onto the *x* axis:

$$F_{\text{c,net}} = 2\pi r_m \gamma_o \sin(\varphi') \sin(\alpha). \tag{4}$$

Note that there is also a vertical force component $F_z = F_z' \cos(\alpha)$, which causes the microdroplet to immerse into the meniscus.

After adding all relevant forces, Newton's second law for the moving droplet is then:

$$\ddot{r} + 6\pi\mu_o (R_1 + \frac{R^{*2}}{r - R_1 - R_2}) \cdot \frac{\dot{r}}{m_1} - k \cdot \frac{2\pi r_m \gamma_o}{m_1} \cdot \sin(\varphi') \sin(\alpha) = 0, \tag{5}$$

where *r* is the center-to-center distance of the two droplets and *m* is the mass of the moving droplet. We use *k* as a fitting parameter to account for errors from the approximations made above and set it as *k* = 0.9 to achieve a good agreement between numerical results and experimental data. To solve eq. (5), the parameters $\varphi'$ and $r_m$ need to be computed from geometrical considerations. We correlate the hemispherical expression for the moving droplet, $y = \sqrt{R_1^2 - x^2}$, and the meniscus profile of the stationary droplet, $h(x) = a \exp(-b(x+r)) + c$, to obtain two intersection points $(x_A, y_A)$ and $(x_R, y_R)$ in the x-z coordinate systems of the moving droplet. We can then write: $r_m \approx (x_A + x_R)^* \sin(\alpha)$. The angle in the air phase $\Theta_a$ is experimentally found to be nearly constant at 150° (see **Fig. S5**). Based on geometry, the relationship $\varphi' = \Theta_{app} - (\pi - \Theta_a)$ is obtained and the angle $\Theta_{app}$ can be computed by differentiating the hemispherical approximation of the moving droplet.[21,46] Finally, we solve the resulting differential equation using Matlab.

**Dependence of droplet velocity on droplet size and oil viscosity**

From Eq. (5), we know the sliding velocity is directly related to the droplet size, the meniscus shape, and the oil viscosity. Here, we first experimentally determine the fitting parameters for the meniscus shape of oil surrounding a stationary droplet with radius $R_2$ = 35 μm. In the following analysis, the size of the stationary droplet is considered to be constant to exclude a change of meniscus size. We compare the velocity obtained numerically by solving eq. (5) to droplets of similar size (*i.e.*, varying $R_1$ with $R_2 \approx$ 35 μm fixed) from the vapor condensation experiments on LIS with different oil viscosities. Fig. 7a shows the relationship between the maximum velocity and the size of moving droplets. Generally, the mathematical solutions are in excellent agreement with the experimental data for GPL 105 and 106, and match reasonably well for GPL 104. We assume that the inaccuracy of capturing meniscus shapes of all individual droplets generally may be the most significant reason causing this error. Furthermore, the assumptions made in developing the analytical model also influence the quality of the match. Specifically, as mentioned in the previous section, the condensed droplets on GPL 102-impregnated surfaces are more mobile compared to those on GPL 106, even for the big droplets (on the order of 100s μm). However, the larger droplets are treated as static in the model, which may also overestimate the relative velocity between the two droplets. Lastly, the experimental results for the maximum velocity are obtained by software auto-tracking the displacement of droplets in videos, which might induce some errors ($\pm$ 40 μm/s) due to vibrations of the setup. Interestingly, we find that the velocity achieves maximum values for $R_1 \approx$ 7-8 μm (excluding peak values from sudden coalescence events) for all viscosities. When the size of moving droplet becomes comparable to that of the stationary droplet (*i.e.*, $R_1 \approx R_2$), the velocity rapidly decreases, as the assumption of $R_1/R_2 \ll 1$ fails and the two droplets equally contribute to the movement. The maximum velocity also approaches zero when the droplet size becomes smaller than $R_1 \approx$ 1.5 μm. We assume these condensed droplets are fully submerged into the oil film and do not move due to the absence of a driving force ($\varphi_A$ and $\varphi_R$ are not defined). This case is similar to the initial phase of condensation (compare to Fig. 3a). Similarly, a droplet of small size might eventually fully submerge into the meniscus of the stationary big droplet. In this process, the resistance force significantly increases and the driving force drops down.

We also examine the relationship of maximum velocity and oil viscosity, as presented in Fig. 7b. Our numerical results are in good agreement with the experimental results, and show the expected decrease in sliding velocity with increasing lubricant viscosity. The numerical results indicate that microdroplets could achieve much higher sliding velocities than those obtained experimentally in this study at even lower lubricant viscosity due to vanishing viscous dissipation.

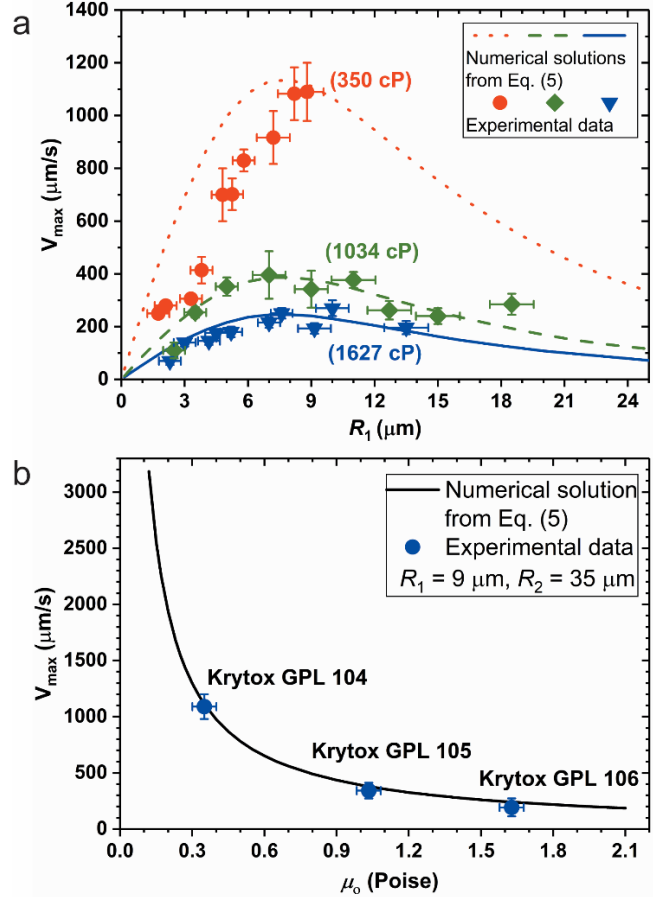

Fig. 7. Experimental and numerical results on the influence of (a) droplet size and (b) lubricant viscosity on the maximum sliding velocity.

## Conclusion, significance, and outlook

In this paper, we investigated the gravity-independent movement of microdroplets (focusing on $D \leq 40$ μm) during water vapor condensation on lubricant-infused surfaces. We experimentally reported a dynamic re-distribution process of the lubricant layer on LIS during vapor condensation, which results in a novel sliding mechanism for condensate microdroplets. In oil-rich regions, the gravity-independent self-propulsion of microdroplets follows a self-similar movement pattern, which is independent of initial lubricant thickness, lubricant viscosity, or droplet size. We successfully modeled this movement based on a lateral force balance. Overlapping lubricant menisci between droplets create an anisotropic oil profile, resulting in unbalanced lateral components of the surface tension force along the droplet – lubricant contact line. We reported that the sliding velocity initially increases due to a relatively big difference of the angle of attack on the two sides of the microdroplet. Then, the velocity sharply decreases as the small droplet approaches the bigger one and submerges into the oil meniscus of the larger droplet. Faster sliding velocities are achieved by decreasing the lubricant viscosity, but the distance and occurrence of sliding events generally decrease due to frequent droplet coalescence and a higher mobility of the larger microdroplets.

Our findings have broad implications on the selection of lubricants and the use of LIS for thermal management and water harvesting systems. This reported novel gravity-independent droplet movement is promising for zero-gravity situations (*e.g.*, in space) or portable devices, where droplet removal and droplet shedding cannot rely on gravity. Furthermore, we expect this capillary-driven droplet motion on LIS to increase sweeping and re-nucleation rates, reduce droplet sizes, and consequently lead to enhanced condensation heat transfer rates as compared to traditional surfaces. We hope this work will be helpful in the selection of a lubricant for LIS to achieve high heat transfer and water collection rates, to enhance our understanding of lubricant depletion and drainage, and to complement similar studies on microdroplet or microparticle movement at air-liquid interfaces.

## Conflicts of interest

We report no conflicts of interest.

## Acknowedgement

The authors would like to thank Jillian Anderson for insightful discussions. The authors acknowledge financial support from the McKelvey School of Engineering at Washington University in St. Louis, and the use of instruments and staff assistance from the Institute of Materials Science and Engineering, the Nano Research Facility (NRF), and the Jens Environmental Analysis Facility. Confocal measurements were performed at the Washington University Center for Cellular Imaging (WUCCI), supported by Washington University School of Medicine, the Children's Discovery Institute of Washington University, St. Louis Children's Hospital (CDI-CORE-2015-505), and the Foundation for Barnes-Jewish Hospital (3770). Confocal data was generated on a Zeiss LSM 880 Airyscan Confocal Microscope which was purchased with support from the Office of Research Infrastructure Programs (ORIP), a part of the National Institutes of Health (NIH) Office of the Director under grant OD021629.